# Here Is Not There: Measuring Entailment-Based Trajectory Similarity for Location-Privacy Protection and Beyond

Zilong Liu[1], Krzysztof Janowicz[1,2], Kitty Currier[2], Meilin Shi[1], Jinmeng Rao[3], Song Gao[3], Ling Cai[4], and Anita Graser[5]

[1]*University of Vienna, Austria*
[2]*University of California, Santa Barbara, USA*
[3]*University of Wisconsin–Madison, USA*
[4]*IBM Research, USA*
[5]*Austrian Institute of Technology, Austria*

While the paths humans take play out in social as well as physical space, measures to describe and compare their trajectories are carried out in abstract, typically Euclidean, space. When these measures are applied to trajectories of actual individuals in an application area, alterations that are inconsequential in abstract space may suddenly become problematic once overlaid with geographical reality. In this work, we present a different view on trajectory similarity by introducing a measure that utilizes logical entailment. This is an *inferential* perspective that considers facts as triple statements deduced from the social and environmental context, in which the travel takes place, and their practical implications. We suggest a formalization of entailment-based trajectory similarity, measured as the overlapping proportion of facts, which are spatial relation statements in our case study. With the proposed measure, we evaluate LSTM-TrajGAN, a privacy-preserving trajectory-generation model. The entailment-based model evaluation reveals potential consequences of disregarding the rich structure of geographical space (e.g., miscalculated insurance risk due to regional shifts in our toy example). Our work highlights the advantage of applying logical entailment to trajectory-similarity reasoning for location-privacy protection and beyond.

## 1 Introduction

Analysing trajectories as a means to study human mobility is not new (Long and Nelson, 2013; Miller et al., 2019). What is new, however, is that such trajectories are now widely available at an unprecedented spatial, temporal, and thematic resolution for individuals (Demšar et al., 2021; Siła-Nowicka et al., 2016; Xu et al., 2022). Over the past few years, this has fueled a rapidly growing marketplace for private companies to sell, buy, and utilize trajectory data for a wide variety of downstream tasks such as insurance assessment, health monitoring, predictive policing, and autonomous vehicle navigation.

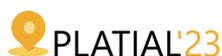

Whether the goal is to mine or learn from trajectories, most research relies on a common set of trajectory similarity measures. Such measures include spatial and spatiotemporal techniques (Magdy et al., 2015). For instance, Fréchet distance (Alt and Godau, 1995) is a common spatial similarity measure, while Dynamic Time Warping (Berndt and Clifford, 1994) falls under the spatiotemporal category. These measures often imply geometry-first thinking (Janowicz et al., 2022) at the expense of place-based context. Another category is semantic-aware measures where a third, thematic component is included, namely attribute labels associated with a trajectory *fix* (Hu et al., 2013). These measures can be applied to semantic trajectories (Parent et al., 2013). Matching or comparing attribute labels is performed using edit distance (Chen and Ng, 2004), Jaccard coefficient (Jaccard, 1901), and so forth.

As with all similarity measures, what matters from a theoretical perspective is that formal properties of trajectory similarity have to be matched to the task at hand. For example, Fréchet distance will be favored over Hausdorff distance (Hausdorff, 1914) if the curvature of the trajectories plays a bigger role in assessing their similarity, while the Jaccard coefficient will be used if thematic knowledge (e.g., place types) is prioritized in the comparison. Therefore, the entire process of finding a suitable trajectory similarity measure requires a detailed understanding of existing measures as well as deep domain knowledge of the application area. Even a small mismatch between measures and application areas can have substantial consequences. This is especially important because results from trajectory analysis are used further downstream as data features for a variety of automated, machine learning-based tasks.

## 2 Why Use Logical Entailment to Measure Trajectory Similarity?

While humans move through physical and social space, trajectory similarity is usually characterized in abstract, typically Euclidean space. This is done largely in order to (1) simplify trajectory representation and similarity computation; (2) develop measures that are domain agnostic, thereby improving their reusability; and (3) reduce the amount of auxiliary data required. However, mobility, from a domain perspective, does *not* happen in abstract space. Oversimplifying trajectory representation and ignoring geographical context can be problematic if mischaracterizing these attributes has real-world consequences.

Put differently, the properties of these measures can have unintended implications when applied to automated decision making regarding real-life geographical space. While some representational and computational simplifications are unavoidable, two major developments encourage us to rethink how to assess similarity. First, we no longer live in a data-poor environment with scarce computational resources. Secondly, the increasingly common usage of data features – such as similarity scores – to set policies, evaluate behaviours, and influence other aspects of social life, has far-reaching and often non-transparent consequences. This is even more important in the case of methods to protect personal location privacy before downstream data mining. To give an intuitive example, a privacy-preserving trajectory-generation model creates synthetic trajectories to protect individual privacy while supporting business analytics, e.g., for insurance companies. To prove that such synthetic trajectories can be used as proxies for real movement data, we need to demonstrate a high similarity between the synthetic and real trajectories. A suitable synthetic trajectory should resemble an individual's real movement patterns without revealing privacy-sensitive details, if not necessary for the application:

$$\text{hd}(t, t') = \max_{(x_i, y_i) \in t} \left\{ \min_{(x'_i, y'_i) \in t'} \text{d}\left((x_i, y_i), (x'_i, y'_i)\right) \right\} \qquad (1)$$

Figure 1 shows two different trajectories, $t$ and $t'$, in geographical space using a place-based representation based on New York City (NYC) community districts from NYC Open Data[1]. The trajectory $t$ is defined as an ordered set of fixes represented by $n$ tuples $\{(x_1, y_1), (x_2, y_2), \ldots, (x_n, y_n)\}$, and its synthetic trajectory $t'$ is defined as $\{(x'_1, y'_1), (x'_2, y'_2), \ldots, (x'_n, y'_n)\}$. Each tuple encodes a location in two-dimensional Euclidean space. While the two trajectories are geometrically similar according to their Hausdorff distance as shown in Equation 1, they pass through a slightly different set of community districts. This small shift may have large consequences in certain contexts. Insurance companies, e.g., may consider regional factors (e.g., neighbourhood crime rates) about where we live and travel when establishing premiums. They may associate a higher crime rate with a higher risk of vandalism and, consequently, charge drivers who visit these regions more for car insurance. Arbitrarily higher or lower fees may result as artifacts of a location privacy-preserving algorithm that alters the type of neighbourhood – with a higher or lower crime rate – a driver is believed to have visited. In NYC, crime rates are reported by community district, which are used here as the areal unit to illustrate an example presented in Section 4. This illustrates how a slight shift in Euclidean space, e.g., 50 meters to the east or west, may nudge a fix across a boundary and into a different geographical region, consequently changing the set of valid *entailed* statements about that fix.

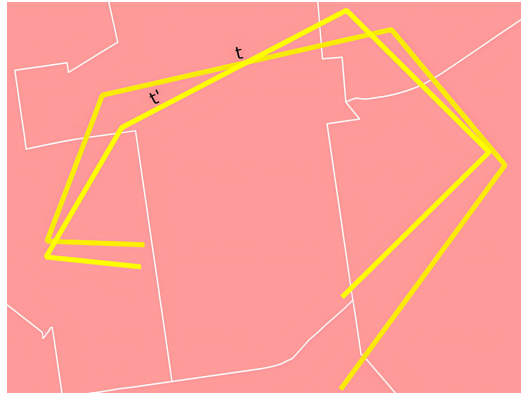

**Figure 1: Two trajectories t and t' overlaid on NYC community districts**

We suggest approaching trajectory analysis by measuring trajectory similarity in terms of entailment, i.e., by the overlap between the sets of all possible inferences deduced from two trajectories. We then consider two trajectories to be similar if they generate the same or a highly overlapping set of entailment-based statements. This perspective is different from most previous work, because it is *inferential* and based on the logical consequence of traveling through actual semantically rich places. Similarity assessment, then, depends on three components: (1) the *terminology* component (TBox), i.e., a set of axioms specifying the classes and relationships within a domain of interest (Figure 2 as an example); (2) the *assertion* component (ABox), i.e., the set of statements about the world, expressed by using the terminology provided; and (3) the *entailment regime*, i.e., the set of rules to be executed over the ABox and TBox to derive inferences.

Next we will introduce how trajectory similarity can be measured via logical entailment. Then, we discuss the measure in the context of privacy-preserving trajectory generation. More concretely, we use the pre-trained version[2] (including trajectory datasets) of LSTM-TrajGAN (Rao et al., 2020) to showcase the impact of considering entailment in model evaluation.

## 3  Similarity as the Overlap of Entailed Statements

Given two trajectories $t$ and $t'$ (from Section 2) and a set of inferential rules, we can derive statements about trajectory fixes from existing contextual knowledge via logical entailment. For the sake of simplicity, in this work we only consider spatial relation statements that can be inferred by a qualitative spatial calculus, e.g., the Region Connection Calculus (Randell et al., 1992). Each statement has a

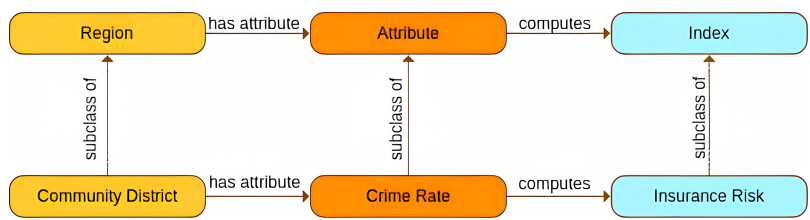

**Figure 2: An example TBox involving NYC community districts in the domain of insurance risk estimation**

triple form consisting of a subject $s$, a predicate $p$, and an object $o$, such as

$$\langle s: \texttt{BronxCommunityDistrict1}, p: \texttt{touches}, o: \texttt{BronxCommunityDistrict3}\rangle.$$

Such a statement and its inverse relation statement, in the example case

$$\langle s: \texttt{BronxCommunityDistrict3}, p: \texttt{touches}, o: \texttt{BronxCommunityDistrict1}\rangle,$$

will be used in measuring similarity if a trajectory fix tf is located in BronxCommunityDistrict1, represented as
$$\langle s: \texttt{tf}, p: \texttt{within}, o: \texttt{BronxCommunityDistrict1}\rangle,$$

In this way, a trajectory crossing BronxCommunityDistrict1 is not counted as entirely dissimilar to a trajectory crossing BronxCommunityDistrict3. Lastly, no equal-relation or disjoint-relation statement is included in the similarity comparison.

Given a ruleset, the set of statements derived from logical entailment over a trajectory $t$ (or $t'$) is the union of all such statements inferred from all its fixes. A Jaccard coefficient applied on discrete sets can be used to compute the entailment-based similarity between $t$ and $t'$ based on the overlapping portion of their statements. We propose the measure

$$J(e, e') = \frac{|e \cap e'|}{|e \cup e'|} \quad (2)$$

where $J(e, e')$ is the entailment-based trajectory similarity, and $e$ and $e'$ represent respective statement sets derived from $t$ and $t'$. We consider statements as *atomic* so that they are not decomposed into subjects, predicates, and objects to compare their similarities. Comparing statement sets is made possible because statements are fully materialized.

## 4  Evaluation Results

Table 1 first shows 10 sample trajectory pairs, along with their directed Hausdorff distance, entailment-based similarity, and insurance risk deviation. From the table, we observe that the Hausdorff distance (in kilometres) is within the range of $[0.78, 1.74]$, which indicates geometric similarity at an acceptable and stable level. However, our entailment-based similarity shows otherwise, i.e., less than 0.5 for all trajectory pairs.

Assume that an insurance company uses trajectory data to assess insurance risk in 2019. To estimate the insurance risk, we use standardized crime rate data from the 2019 NYC community district profile[3]. We then average crime rates of all community districts that a person crossed to estimate the corresponding insurance risk. The last column of Table 1 shows the deviation of insurance risk if using synthetic trajectories instead of real ones. While the exact deviation values will depend on the context of the parameters being characterized – in our case, using crime rate as a proxy for insurance risk – this toy example demonstrates that geometrically similar trajectories can give rise to unanticipated deviations in insurance risk. Therefore, using synthetic trajectories derived from LSTM-TrajGAN may lead to unreasonable insurance rates, caused by small shifts of trajectory fixes from one region to another. Lastly, we find that among 884 synthetic trajectories there are 417 trajectories containing fixes outside the community district data layer. Lacking crime rate data for those fixes, those synthetic trajectories cannot be evaluated adequately against the others.

Table 1: **Ground-truth and synthetic trajectories.** The table includes a comparison between Hausdorff distance and entailment-based similarity of the first 10 pairs of ground-truth and synthetic trajectories. The toy example *insurance risk deviation* shows the difference in risk – artificially higher or artificially lower – calculated by a hypothetical insurance company using the synthetic trajectory rather than the ground-truth trajectory.

| Trajectory pair ID | Hausdorff distance [km] | Entailment-based similarity | Insurance risk deviation |
|---:|---:|---:|---:|
| 126 | 1.38 | 0.38 | 3.06 |
| 131 | 1.74 | 0.38 | 2.19 |
| 133 | 1.31 | 0.39 | 0.74 |
| 135 | 1.63 | 0.43 | 3.92 |
| 143 | 0.78 | 0.47 | 0.11 |
| 144 | 0.83 | 0.37 | -0.05 |
| 170 | 1.17 | 0.45 | 1.43 |
| 177 | 0.63 | 0.46 | 0.22 |
| 286 | 1.61 | 0.43 | 0.79 |
| 289 | 1.13 | 0.47 | 3.42 |

## 5 Here Is Not There

While current methods ignore whether a user's fix is shifted into one neighbourhood or another, using entailment-based measures gives us a way to incorporate a *platial* perspective, or the idea that where a trajectory takes place in physical reality matters. While our example in this work is purely topological for simplicity, semantically rich assertions (e.g., social vulnerability) available in all these places could be included in Equation 2 without any modification.

The concept of trajectory similarity based on logical entailment has implications for the dilemma in protecting an individual's location privacy, i.e., how to protect privacy while preserving the utility of their data. In this work, we demonstrate that the importance of utility cannot be overlooked because human-centric downstream tasks are concerned with place-based context, which requires more attention in future work on location-privacy protection. As choosing the kind of places that matters to a downstream task often introduces a modifiable areal unit problem (Fotheringham and Wong, 1991), the usage of the TBox in measuring entailment-based similarity helps deal with it by pre-defining the relationship between places and tasks.

When computing entailment-based trajectory similarity, the statements considered for similarity reasoning depend on the actual rule set used to infer them. An example of such inferential rules is the transitivity of spatial containment. The more expressive the entailment regime used, the more statements will be entailed and materialized, influencing both the similarity score and the computational complexity of our measure.

## 6 Conclusions

In this work, we present a logical-entailment view of trajectory similarity. This inferential perspective leads us to compare facts that can be deduced from a place-based context. To compute entailment-based similarity, we measure the overlap of ABox statements and use spatial relation statements as an example. By evaluating the utility of a privacy-preserving trajectory-generation model, LSTM-TrajGAN, we demonstrate how the proposed similarity measure can reveal what is hidden by geometry-based measures operating on an abstract Cartesian plane, and how its usage increases awareness of the application area. In the future, we plan to use large-scale mobility datasets to determine the relevance of a statement to a trajectory based on TF–IDF (Salton and Buckley, 1988) and create local knowledge graphs to support inferences with the help of geo-enrichment services. Also, while in our study we use qualitative spatial calculi, we will explore how to generalize the proposed measure to other calculi, e.g., place-type hierarchies to compute place type-based trajectory similarity, or Allen's interval algebra (Allen, 1983) for temporal reasoning to compute order-constrained trajectory similarity. To sum up, our work demonstrates that measures that fail to take into account actual places and their social context may lead to unintended consequences in an era of massive utilization of machine learning-based recommendations.

### Notes

1. https://opendata.cityofnewyork.us
2. https://github.com/GeoDS/LSTM-TrajGAN
3. https://communityprofiles.planning.nyc.gov

**Acknowledgements**

This work is partially funded by EU's Horizon Europe research and innovation programme under Grant No 101093051 EMERALDS.

**ORCID**

Zilong Liu      https://orcid.org/0000-0002-7699-3366
Kitty Currier      https://orcid.org/0000-0001-8020-8888
Meilin Shi      https://orcid.org/0000-0001-6039-7810
Jinmeng Rao      https://orcid.org/0000-0003-2370-5129
Song Gao      https://orcid.org/0000-0003-4359-6302
Ling Cai      https://orcid.org/0000-0001-7106-4907
Anita Graser      https://orcid.org/0000-0001-5361-2885